\documentclass{iaus}
\usepackage{graphicx}

\title[The rotation of very low-mass stars and brown dwarfs] 
{The rotation of very low-mass stars and brown dwarfs}

\author[Jochen Eisl\"offel and  Alexander Scholz]   
{Jochen Eisl\"offel$^1$ \and 
 Alexander Scholz$^2$}

\affiliation{$^1$Th\"uringer Landessternwarte, Sternwarte 5, D-07778
  Tautenburg, Germany\\[\affilskip]
$^2$SUPA, School of Physics \& Astronomy, University of St. Andrews, North
  Haugh, St. Andrews, Fife KY16 9SS, United Kingdom}

\pubyear{2007}
\volume{IAUS243}  
\pagerange{1--1}
\date{?? and in revised form ??}
\jname{Proceedings Star-disk interaction in young stars}
\editors{J. Bouvier Editor}
\begin{document}

\maketitle

\begin{abstract}
The evolution of angular momentum is a key to our understanding of star
formation and stellar evolution. The rotational evolution of solar-mass stars
is mostly controlled by magnetic interaction with the circumstellar disc and
angular momentum loss through stellar winds. Major differences in the internal
structure of very low-mass stars and brown dwarfs -- they are believed to be
fully convective throughout their lives, and thus should not operate a
solar-type dynamo -- may lead to major differences in the rotation and
activity of these objects.

Here, we report on observational studies to understand the rotational
evolution of the very low-mass stars and brown dwarfs. 

\end{abstract}

\firstsection 
\section{Introduction}
The study of stellar rotation during the pre-main sequence (PMS) and the main
sequence (MS) phase has provided us with many new insights into their 
formation and evolution (\cite[Bodenheimer 1995]{b95}, \cite[Bouvier et al. 
1997]{bfa97}, \cite[Stassun \& Terndrup 2003]{st03}, \cite[Herbst et al. 
2007]{hems07}). The so-called angular momentum problem of star formation asks
how the specific angular momentum (angular momentum / mass) of dense molecular
cloud cores, from which low-mass stars form, gets reduced by 5 -- 6 orders of
magnitude compared to what is left in solar-type stars on the zero-age main
sequence (ZAMS). An additional 1 -- 2 orders of magnitude is lost from the
ZAMS to the age of the Sun ($\sim$ 5 Gyr).

At the early formation stages, magnetic torques between the collapsing cloud
core and the surrounding interstellar medium, the deposition of a large amount
of angular momentum into the orbital motion of a circumstellar disc, a
planetary system, and/or a binary star system play important roles. It is
known that solar-mass stars already in their T\,Tauri phase rotate
slowly, although they are still accreting matter from their disc. Magnetic
coupling between the star and its circumstellar disc, and the consequent
removal of angular momentum in a highly collimated bipolar jet are considered
to be the agent for this rotational braking (e.g., \cite[Camenzind 1990]{c90}, 
\cite[K{\"o}nigl 1991]{k91}, \cite[Shu et al. 1994]{snowrl94}, 
\cite[Matt \& Pudritz 2005]{mp05}). 
After the cessation of accretion and the following dispersal of the disc this
braking mechanism obviously comes to an end, and the rotation is observed to
accelerate as the PMS stars contract towards the ZAMS. On the main sequence
then, the rotation rates of solar-mass stars  decrease again, because now
angular momentum loss through stellar winds takes over as the dominant
process.

The rotation of stars can either be measured spectroscopically from the
line-broadening of photospheric spectral lines, or it can be derived from
periodic variability in the light curves of stars seen in photometric time
series observations. While the spectroscopic method suffers from projection
effects -- the inclination angle of the rotation axis with respect to our line
of sight is unknown -- the photometric method allows us to determine the
rotation period with high precision and independent of inclination angle.

Whereas a large amount of rotation periods are now available in the literature
for low-mass stars in clusters younger than about 3 Myr (ONC: \cite[Herbst 
et al. 2001]{herbst01}, \cite[Herbst et al. 2002]{hbm02}; NGC2264: 
\cite[Lamm 2003]{l03}, \cite[Lamm et al. 2004]{lbm04}), and most recently 
in the much older M34: \cite{iahibchm06} and NGC2516: \cite{ihahbcmb07}, 
not much is known about the early evolution of very
low-mass (VLM) stars and brown dwarfs up to the age of a few Gyr, when they
are found as field object in the solar neighbourhood (\cite[Bailer-Jones 
\& Mundt 2001]{bm01}, \cite[Clarke et al. 2002]{ctc02}). This has lead us to
initiate a programme to study the rotation periods of the VLM stars and brown
dwarfs, and to compare them to solar-mass stars as well as to evolutionary
models. For our monitoring programme we decided to follow the photometric time
series approach to obtain precise rotation periods. 

In this text, we first present our results on rotation rates and variability
of the sources in Section 2. The observed rotation rates are then compared
to various models of rotational evolution in Section 3. In Section 4 we
discuss some first attempts to characterise the spots on VLM objects, followed
by some comments about accretion and time variability in these
objects in Section 5. Finally, Section 6 contains our conclusions.

\section{Rotation and variability of VLM objects}

In the course of our ongoing monitoring programme, we have so far created a
database of rotation periods for 23 VLM objects in the cluster around sigma
Ori (\cite[Scholz \& Eisl{\"o}ffel 2004a]{se04a}), for 30 in the field around
epsilon Ori (\cite[Scholz \& Eisl{\"o}ffel 2005]{se05}), which are belonging
to the Ori\,OBIb association, and for 9 objects in the Pleiades open cluster
(\cite[Scholz \& Eisl{\"o}ffel 2004b]{se04b}). At ages of about 3, 5, and
125\,Myr these three groups of VLM objects form an age sequence that allows us
some insights into a relevant part of their PMS evolution. 

In general, the observed periodic variability in the light curves of our VLM
targets is attributed to surface features, which are asymmetrically
distributed on the surface and are co-rotating with the objects. Such surface
features may arise from dust condensations in the form of ``clouds'',
or from magnetic activity in the form of cool ``spots''. 
Because of their youth all our objects have surface temperatures T$_{\rm eff}$
$>$ 2700\,K (\cite[Baraffe et al. 1998]{baraffe98}), which corresponds to
spectral types earlier than M8. These temperatures are higher than the dust
condensation limits, so that we are most likely observing cool, magnetically
induced spots.

\begin{figure}[t]
  \begin{center}
    \resizebox{120mm}{!}{\includegraphics[angle=-90]{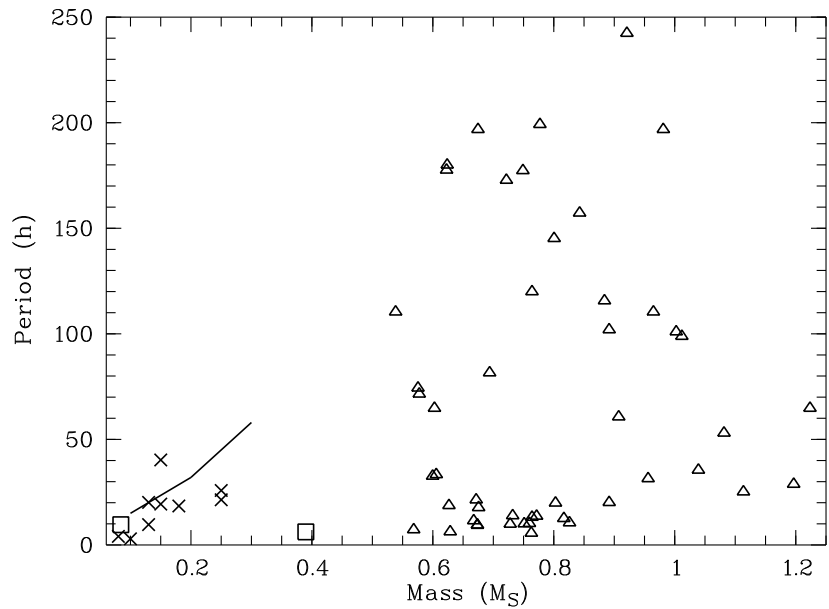}}
  \end{center}
\caption{Rotation periods versus mass in the Pleiades. Our rotation periods
  for VLM objects are shown as crosses. Triangles mark the periods for 
solar-mass stars from the Open Cluster Database. The two squares show 
periods from Terndrup et al. (1999) The solid line marks the upper 
limit to the observed $v \sin i$ values of Terndrup et al. (2000).}
\end{figure}

It is interesting to compare the mass dependence of the rotation periods for
the VLM and solar-mass objects. For the Pleiades, we find that their period 
distributions are different. While periods of up to ten days are known for 
the solar-mass objects, Fig.\,1 shows that among the VLM objects we are 
lacking members with rotation periods of more than about two days. 
Although our photometric time series extends over a time span of 18 days, 
slow rotators might have been missed among the VLM objects, if their spot
patterns evolved on a much shorter time scale, or if they did not show any
significant spots. 

These possibilities can be checked by converting the spectroscopically derived
lower limits for rotational velocities from \cite{terndrup99} and references
therein into upper limits for the rotation periods of the VLM objects using
the radii from the models by \cite{chabrier97}. Such spectroscopically derived
rotational velocities should not be affected by the evolution of spot patterns
on the objects. The derived upper period limits are shown in Fig.\,1 as a
solid line. With a single exception, all our data points fall below this line.
Hence, they are in good agreement with the spectroscopic rotation
velocities. Both complementary data sets indicate the absence of slow rotators
among the VLM objects. Looking at them in detail, they even show a trend
towards faster rotation even within the VLM regime from higher to lower
masses. Such a trend is also seen in our data of the epsilon Ori cluster, and
in the Orion Nebula Cluster data by \cite{herbst01}.

\section{Rotational evolution of VLM objects}

We now want to combine the periods for all three clusters that we observed,
namely sigma Ori (\cite[Scholz \& Eisl{\"o}ffel 2004a]{se04a}), epsilon Ori
(\cite[Scholz \& Eisl{\"o}ffel 2005]{se05}), and the Pleiades (\cite[Scholz \&
  Eisl{\"o}ffel 2004b]{se04b}), to try to reproduce their period distributions
with simple models. These models should include the essential physics of star
formation and evolution that we described in Section 1. A practical way of
doing this is to project the period distribution for sigma Ori forward in time
and then compare the model predictions with our observations for epsilon Ori
and the Pleiades. 

\begin{figure}[ht]
  \begin{center}
    \resizebox{120mm}{!}{\includegraphics[angle=-90]{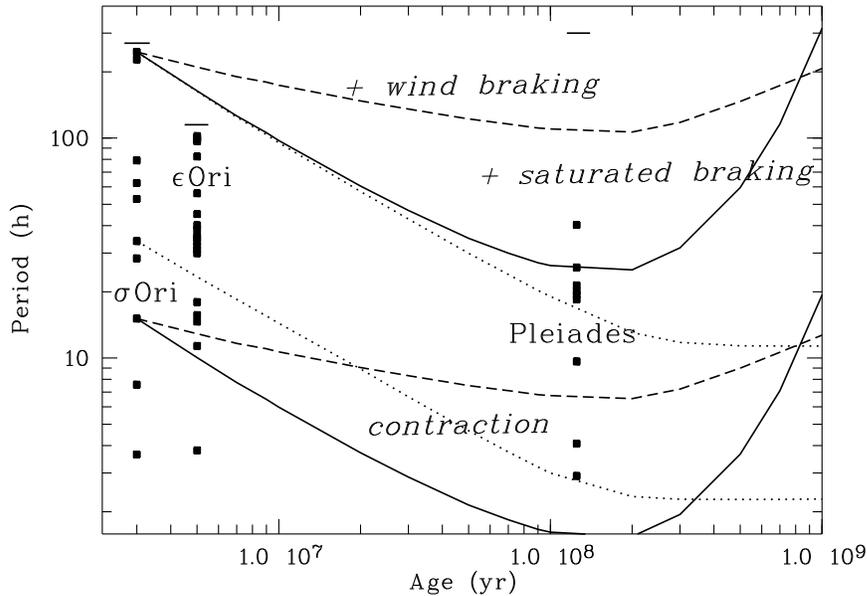}}
  \end{center}
\caption{Rotational evolution of VLM objects. The dotted lines show the
  evolution of the rotation periods for a couple of objects for a model of 
  hydrostatic contraction only. A model with additional Skumanich type 
  braking through stellar winds is shown as dashed lines, and models that use
  a saturated wind braking instead are shown as solid lines.}
\end{figure}

It is obvious, that the hydrostatic contraction of the newly formed VLM
objects has to be taken into account in a first step. Changes in their
internal structure may be negligible for these fully convective objects
(\cite[Sills et al. 2000]{spt00}). Then, their rotation periods should evolve
from the initial rotation period at the age of sigma Ori strictly following 
the evolution of
the radii (which were taken from the models by \cite[Chabrier \& Baraffe 
1997]{chabrier97}). 
Therefore, the rotation accelerates, and should only level out for ages 
older than the Pleiades, when the objects have settled (dotted lines in 
Fig.\,2). This model, however, is obviously in conflict with the observed
Pleiades rotation periods. Half of the sigma Ori objects would get accelerated
to rotation periods below the fastest ones observed in the Pleiades of about
3\,h. Furthermore, even the slowest rotators in sigma Ori would get spun up to
velocities much faster than the slower rotators in the Pleiades, which would
remain unexplained then. This teaches us that significant rotational braking
must occur until the objects reach the age of the Pleiades, because it is
clear that the sigma Ori VLM objects will undergo a significant contraction
process. 

In a second model we now add a Skumanich type braking through stellar
winds, as it is seen in solar-type MS stars (\cite[Skumanich 1972]{s72}). This
wind braking increases the rotation periods $\sim$ $t^{1/2}$, as shown by the
dashed lines in Fig.\,2. However, following this model some of the sigma Ori 
slow rotators would get braked far to
strongly. For a Skumanich type wind braking they would become clearly slower
rotators than are observed in the Pleiades (see also above). A possible
solution to this problem is to assume that even the slowest 
sigma Ori rotators seem to rotate so fast, that they are beyond the 
saturation limit of stellar winds (\cite[Chaboyer et al. 1995]{cdp95}, 
\cite[Terndrup et al. 2000]{tsp00}, \cite[Barnes 2003]{b03}). 
In this saturated regime, angular momentum loss is assumed to depend only
linearly on angular rotational velocity, and therefore rotation periods
increase exponentially with time. The solid lines in Fig.\,2 show a model of
contraction and saturated wind braking. The period evolution of this
model is the most consistent with our data.

\begin{figure}[th]
  \begin{center}
    \resizebox{120mm}{!}{\includegraphics[angle=-90]{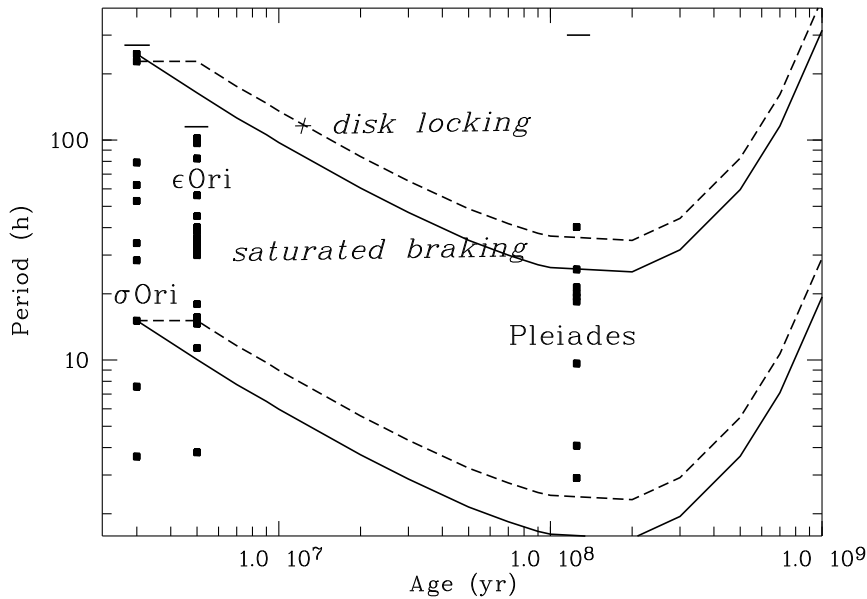}}
  \end{center}
\caption{Rotational evolution of VLM objects. The evolution of the rotation 
periods for a couple of objects for a model with hydrostatic contraction and 
saturated wind braking are shown as solid lines, as in Fig.\,2, while a model
with added disc-locking up to an age of 5\,Myr is shown as dashed lines.}
\end{figure}

It is interesting to explore if disc-locking -- as an angular momentum
regulator active only at very young ages -- may also play a role for the
evolution of rotation periods. The sigma Ori cluster would be young enough for
this process to play a role, and indeed we found evidence that some of our
objects in this cluster may possess an accretion disc. 
Therefore, we investigate a scenario in which disc-locking is active for an
age up to 5\,Myr, typical for the occurrence of accretion discs in solar-mass
stars. During this time rotation periods would remain constant. This
disc-locking scenario we combined with the saturated wind braking, with an
adapted spin-down time scale. In Fig.\,3 dashed lines are shown for two
objects following this model, together with the pure saturated wind braking
model discussed above (the solid lines, as in Fig.\,2). The period evolution
for both models is nearly indistinguishable. It turns out that from our
currently available rotation periods for these three clusters, strong
constraints for or against disc-locking on VLM objects cannot yet be placed.

\section{The spots of VLM objects}

\begin{figure}[th]
\resizebox{120mm}{!}{\includegraphics[angle=-90]{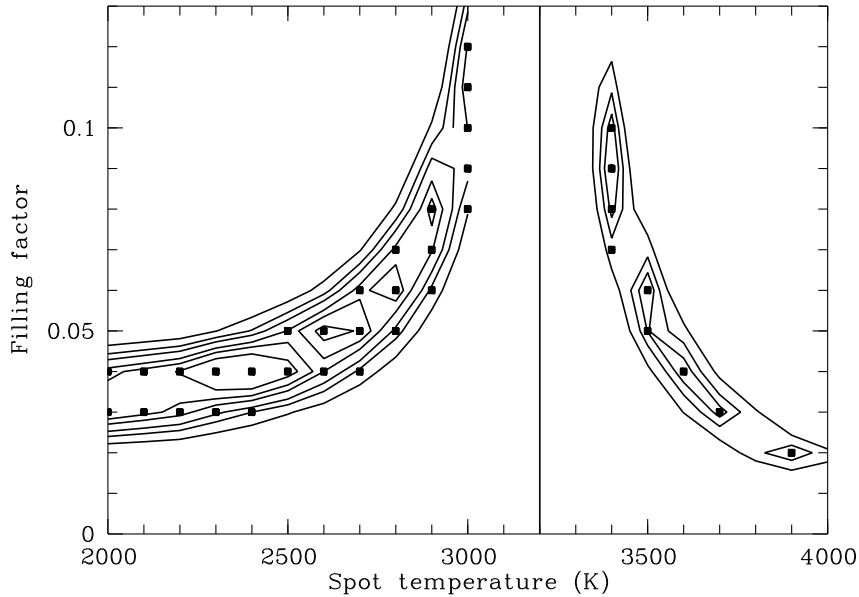}}
\caption{Contour plot for the $\chi^2$ values from the comparison between
  observed and simulated spot amplitudes in the Pleiades VLM star BPL\,129. 
  Contour lines start at $\chi^2=$3.0 and
  are plotted for $\chi^2=$3.0, 2.0, 1.5, 1.0, 0.75, 0.5, indicating
  increasing quality of the fit. Filled squares show all combinations of spot
  temperature and filling factor which would produce amplitudes within the
  error bars of our observations. The vertical solid line indicates the
  photospheric temperature of BPL\,129. Note that the hot spot solutions on the
  right side never reach to the 4th contour line, and $\chi^2$ is larger than
  0.9 everywhere. They are thus significantly worse than the cool spot
  solutions on  the left side. (One data point with $\chi^2 = 0.92$ at $T_S =
  3300$\,K and $f = 19$\% is not shown in the figure.)}
\end{figure}   

Not much is known about the properties of the spots on VLM stars and brown
dwarfs. First clues on 
the physical properties of the spots may be obtained from multi-filter time
series observations. In principle, they allow us to measure the variation
amplitudes in the light curves at various wavelengths, and from this
information to derive the (asymmetric part of the) spot filling factor and the
temperature difference between the spots and the average atmosphere --
although this method is not capable of delivering unique solutions. 

Therefore, in parallel to a photometric time series campaign of the Pleiades
in the I-band, we measures in the J- and H-band simultaneously on a second 
telescope (\cite[Scholz et al. 2005]{sef05}). Only one VLM star (BPL\,129) 
showed a period in all three
wavelength bands at a signal-to-noise high enough so that we could derive spot
properties. For several other Pleiades VLM stars and brown dwarfs only limits
could be placed. 

The best agreement between the observations and a one-spot model is reached
for a cool spot with a temperature 18 to 31\% below the average photospheric
temperature and a filling factor of 4 -- 5\% (see Fig.\,4). These results
indicate that spots on VLM stars may have a similar temperature
contrast between  spot and average atmosphere, but a rather low spot filling
factor compared to solar-mass stars. This might be a consequence of a change
in the dynamo from a solar-type shell dynamo to a small-scale turbulent dynamo
in these fully convective stars.

\section{Accretion and time variability}

\begin{figure}[th]
  \begin{center}
    \resizebox{120mm}{!}{\includegraphics[angle=-90]{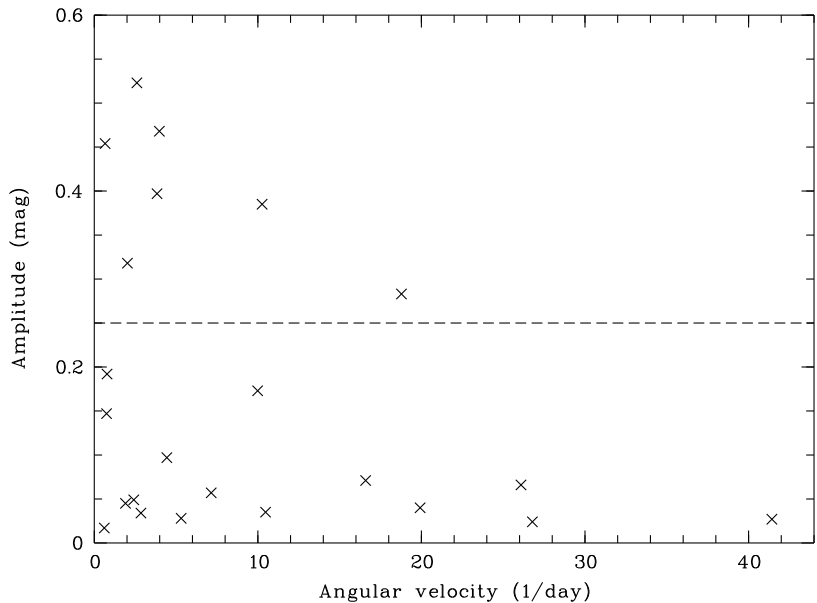}}
  \end{center}
\caption{Angular velocity versus amplitude of VLM stars and brown dwarfs in
  the sigma Ori cluster. The dashed line delineates the 
separation between low-amplitude and high-amplitude objects. The
high-amplitude objects are mostly active accretors, and on average rotate 
slower that the non-accreting low-amplitude objects.}
\end{figure}

In the course of our data analysis we noted that a few of the VLM objects in
the two Orion regions show large amplitudes of up to 0.6\,mag (see
Fig.\,5). These variations are, however, much of an irregular
character. Because of the large amplitudes, it is most likely that they result
from hot spots originating from accretion of circumstellar disc matter onto
the object surface (see also \cite[Fern\'andez \& Eiroa 1996]{fe96}). 
Emission lines in H$\alpha$, the
far-red Calcium triplet, and -- in some cases -- even forbidden emission lines
of [OI]$\lambda\lambda$6300,6363 and [SII]]$\lambda\lambda$6716,6731 are seen
in optical spectra that we obtained of some of these objects in sigma Ori.
These spectra thus show signatures typical of accretion, much like those of
classical T\,Tauri stars. In addition, in a near-infrared
colour-colour-diagram the high-amplitude variables mostly lie in the
reddening path or even red-ward of it, thus showing near-infrared excess
emission that is usually taken as evidence for a circumstellar disc.
With their photometric variability, spectral accretion signatures, and
indications for near-infrared excess emission from discs appear to be the
low-mass and substellar counterparts to solar-mass T\,Tauri stars.

It is interesting that the high-amplitude T\,Tauri analogs on average 
are slower rotators than their low-amplitude non-accreting siblings.
A similar tendency that brown dwarfs with discs seem to rotate more slowly is
also seen in spectroscopic measurements of v\,sin\,i by \cite{mjb05}. 
It thus seems that even in the substellar regime a connection between 
accretion and rotation exist, possibly implying rotational braking due to
interaction between object and disc (\cite[Scholz \& 
Eisl{\"o}ffel 2004a]{se04a}).

\section{Conclusions}

We report results from our ongoing photometric monitoring of VLM objects 
in the clusters around sigma Ori, epsilon Ori, and the Pleiades, and our 
first attempts to model their rotational evolution. 

It is very likely that the observed periodic variability of many VLM objects 
originates from magnetically induced cool spots on the surfaces of the 
objects. In particular in the Pleiades, variation amplitudes in VLM objects
indicate either less asymmetric spot distribution, smaller relative spotted
area, or lower contrast between spots and average photosphere than in
solar-mass stars. VLM objects show shorter rotation periods with decreasing
mass. This effect is observed already at the youngest ages, and therefore
should have its origin in the earliest phases of their evolution. 

Combining the rotation periods for all our objects, we find that their
evolution does not follow hydrostatic contraction alone. Some kind of
braking mechanism, e.g. wind braking similar to the one observed in solar-mass
stars, is required as well. Such a wind braking is intimately connected to
stellar activity and magnetic dynamo action (\cite[Schatzman
  1962]{schatzman62}). Nonetheless, since all the 
investigated VLM objects are expected to be fully convective, they 
should not be able to sustain a solar-type large-scale dynamo, which is at the
heart of the Skumanich type angular momentum loss of solar-mass stars. In
fact, our modelling shows that such a Skumanich type wind braking cannot
explain our data, while saturated angular momentum loss following an
exponential braking law can. This and the observed small photometric
amplitudes may advocate a change in the magnetic field generation in the VLM
regime. The exact type of dynamo operating in VLM objects is unclear. In
principle, observations of rotation bear great potential to distinguish
between the various scenarios for such dynamos (e.g., \cite[Durney et al. 
1993]{ddr93}, \cite[Chabrier \& K\"uker 2006]{ck06}). However, consistent 
theoretical models that provide
predictions for rotational braking in the very low-mass regime and thus
rigorous testing against the observations, are not yet available.

\begin{acknowledgments}
This work was partially supported by Deutsche Forschungsgemeinschaft (DFG) 
grants Ei 409/11-1 and 11-2, and by the European Community's Marie Curie 
Actions--Human Resource and Mobility within the JETSET (Jet Simulations, 
Experiments and Theories) network under contract MRTN-CT-2004 005592.
\end{acknowledgments}

\end{document}